\documentclass[12pt]{article}
\usepackage{graphicx}

\def\beq{\begin{equation}}
\def\eeq#1{\label{#1}\end{equation}}
\def\eeqn{\end{equation}}
\def\beqa{\begin{eqnarray}}
\def\eeqa#1{\label{#1}\end{eqnarray}}
\def\eeqan{\end{eqnarray}}

\def\ee{e^+e^-}

\def\newsection#1{

 \bigskip

\noindent {\bf #1}

\medskip

}

\begin{document}
\begin{center}
{\large {\bf Statement from the Americas Linear Collider Committee \\
to the P5 subpanel}}
\end{center}

\bigskip
\begin{center}
J.A.~Bagger$^{1,2}$,  
S.~Belomestnykh$^{3}$,     
P.C.~Bhat$^{3}$,     
J.E.~Brau$^{4}$, 
M.~Demarteau$^{5}$, 
D.~Denisov$^{6}$, 
S.~Gori$^{7}$, 
P.D.~Grannis$^{8}$, 
T.~Junginger$^{9,10}$,  
A.J.~Lankford$^{11}$,
M.~Liepe$^{12}$,
T.W.~Markiewicz$^{13}$,
H.E.~Montgomery$^{14}$, 
M.~Perelstein$^{12}$,
M.E.~Peskin$^{13}$, 
J.~Strube$^{15,4}$, 
A.P.~White$^{16}$,
G.W.~Wilson$^{17}$
\end{center}

\begin{center} 
$^{1}$Johns~Hopkins~University,
$^{2}$University~of~British~Columbia,
$^{3}$Fermi~National~Accelerator~Laboratory,
$^{4}$University~of~Oregon,
$^{5}$Oak~Ridge~National~Laboratory, 
$^{6}$Brookhaven~National~Laboratory,
$^{7}$University~of~California~Santa~Cruz,
$^{8}$Stony~Brook~University,
$^{9}$University~of~Victoria,
$^{10}$TRIUMF,
$^{11}$University~of~California~Irvine, 
$^{12}$Cornell University, 
$^{13}$SLAC~National~Accelerator~Laboratory,
$^{14}$Thomas~Jefferson~National~Accelerator~Laboratory,
$^{15}$Pacific~Northwest~National~Laboratory,
$^{16}$University~of~Texas~Arlington,
$^{17}$University~of~Kansas
\end{center}


\begin{abstract}
This statement from the Americas Linear Collider Committee to the P5 subpanel has three purposes. It presents a brief summary of the case for an $\ee$ Higgs factory that has emerged from Snowmass 2021. 
It highlights the special virtues of the ILC that are shared with other linear colliders but not with circular colliders. 
Finally, it calls attention to the resources available in an ILC White Paper submitted for for Snowmass 2021.
The ALCC urges P5 to move the Higgs factory forward as a global project by assigning the idea of an $\ee$ Higgs factory high priority, initiating a global discussion of the technology choice and cost sharing, and offering the option of siting the Higgs factory in the U.S.
    
\end{abstract}

\bigskip

We are writing to convey the input of the Americas Linear Collider
Committee (ALCC) to the P5 deliberations.  
The ALCC provides the interface between the international
organization
of the International Linear Collider (ILC) and laboratories, universities, and
governmental agencies in the U.S. and elsewhere in the Americas.  The
members of the ALCC have devoted considerable time and effort over
many years to the realization of the ILC because we feel that this
project is essential to the progress of particle physics.

The Snowmass 2021 process has led to a great deal of
information-gathering and community discussion on the future of the
Energy Frontier in general and on $\ee$ Higgs boson factories in
particular.  This has led to a refined understanding of the
case for an $\ee$ collider as the next global project in particle 
physics.  This letter then has three purposes.  First, we 
present a summary of the case for a Higgs factory that has 
emerged from Snowmass 2021.  Second, we highlight the special virtues
of the ILC that are shared with other linear colliders, but not with
circular
collider Higgs factory proposals.  Finally,  we  call your attention to the 
resources that we have made available to you in the ILC White Paper for
Snowmass (arXiv:2203.07622), which our committee has put together in
collaboration with the 
ILC International Development Team
established by ICFA
and with the ILC
community.  That report provides a detailed justification for all 
aspects of the summary statement that we provide here.
We will separately provide additional brief notes on budget and schedule of accelerator and experiment R\&D and engineering design.

\newsection{The case for a Higgs factory}

Why is it so important to study the Higgs boson?
It is well understood that the Higgs boson has a central role in the
Standard Model.  But what must be emphasized even more is that the
Higgs boson is centrally responsible for the {\it mysteries} of the Standard
Model.  The framework of the Standard Model is provided by its
particle content and gauge symmetry, but the physics of the model is
controlled by numerous parameters whose origin we do not
understand. The key to the weak interactions is the 
spontaneous breaking of the gauge symmetry (the Higgs mechanism). But  
whether or not this occurs depends on the Higgs mass parameter 
and self-coupling.   The
masses of the $W$ and $Z$ bosons, which set the scale of weak
interactions, also depend on these values. Physics models that seek to
explain these values invoke new particles, from additional bosons of a
larger Higgs sector to concepts such as supersymmetry and
Higgs field compositeness.  The very existence of electroweak symmetry
breaking suggests that there are new interactions at short distances
waiting to be discovered. 

Our lack of understanding
of the Higgs boson is actually a crucial issue for all of the
frontiers of particle physics.  Flavor physicists seek to explain  the
masses and mixings of quarks and leptons.   But in the Standard Model,
these arise from the fermion-Higgs Yukawa couplings.  CP violation
also appears uniquely in the Yukawa couplings.  Models of flavor are
built  using the ingredients of the Higgs sector models discussed in the
previous paragraph.  If we have the wrong picture there, we cannot
build a correct model of flavor.  Similarly, neutrino masses and
mixings require electroweak symmetry breaking and so must also rely on
our understanding of the Higgs field.  In the cosmic frontier, it is
possible to build models of dark matter that are independent of any
other particle sector.  But the most studied models of dark matter
either use the new interactions responsible for electroweak symmetry
breaking (as with  supersymmetric dark matter candidates) or describe
the dark matter particle itself as a Higgs-like particle (as with
the axion).   For all of these questions, there is no full
understanding until we understand the Higgs field.

Since the values of the Standard Model parameters are now well known,
we can make precise predictions within this model and search for
anomalies that would signal new physics.  We are now carrying out such
searches at low energies with precision measurements, in rare weak
interaction particle decays, and in particle searches at the high
energies offered by the LHC.   Oddly, the one place where we are not
yet placing significant constraints on new physics is in  the study of the
Higgs boson itself.  Explicit models of new physics predict deviations
in the Higgs boson properties at the 10\% level and below.  Currently,
the LHC measurements of Higgs boson processes, however technically
impressive, are not yet in the game of searching for new physics.  By
the end of the HL-LHC, the ATLAS and CMS experiments expect to
constrain Higgs boson couplings at the few-percent level.   This may
be sufficient to observe deviations from the Standard Model, but it is 
likely not sufficient to prove that the observed deviations are real. 
 For this, we will need a different
collider that is  better adapted to achieving high
precision while controlling systematic errors. 

The ALCC and its international partners have given considerable
thought to the issue of observing beyond-Standard-Model effects in the
Higgs boson properties.  We believe that experiments with $e^+e^-$ colliders are
necessary.  At hadron colliders, even the simple observation of Higgs
boson processes is complex and requires a sophisticated analysis. In
$e^+e^-$, this starting point is straightforward to achieve.  From this
point, one can go forward with increasing sophistication to
 measure the rates and distributions for Higgs boson processes
with detectors of higher capability than those possible at the LHC,
and with more independent observables to provide cross-checks on the level of
systematic uncertainties.  We anticipate measuring all of the leading
Higgs boson couplings to better than 1\% precision and the couplings
to $W$ and $Z$ to parts-per-mil precision.  These coupling determinations
will be highly robust, in particular, constraining the Higgs boson
width  in a
model-independent way, even in the presence of exotic decay modes.
Thus, we expect to measure the full pattern
of Higgs boson decays, exposing deviations from the Standard Model
predictions 
 with high significance.  This pattern in turn will  give clues to the ultimate
 origin of electroweak symmetry breaking and the principles
 underlying the Higgs field.

At the same time, we will systematically improve our knowledge of
precision electroweak observables.  Higgs factories will dramatically
improve our knowledge of the nonlinear couplings of the $W$ and $Z$
bosons.  And, Higgs factories will be able, for the first time, to
carry out
precision measurements of the electroweak couplings of the top quark.
Many models of new physics that attempt to explain the origin and
importance of the large size of the top quark mass predict deviations
from the Standard Model
in these couplings and in the coupling of the top quark to the Higgs field.

A worldwide consensus has developed that these measurements are
centrally important for particle physics, and that the next global
accelerator should be an $e^+e^-$ Higgs factory. This conclusion is 
front and center  in the report from the most recent update of the European Strategy
for Particle Physics and in the Snowmass 2021 Energy Frontier report. 
 Such a collider is
not inexpensive, but its cost is well within the resources of the 
global particle physics community.

There is also a cost to not  building a Higgs factory.
Today, the technologies for $\ee$
Higgs factories are the only accelerator technologies  that are
fully mature, allowing a new collider to be built and operating before 2040.
Without a Higgs factory on this time scale, we risk a long period
with no operating collider in the world, and perhaps the end of
collider physics altogether.

  But, we are not making progress toward this goal.
 The most mature proposal, the International Linear Collider in Japan,
appears to be 
stalled at the level of Japanese government.  CERN has proposed the
circular collider  FCC-ee but by itself does not appear to have the  resources
to realize this project until the end of the 2040's. FCC-ee is
necessarily
entangled with the prospects for
FCC-hh, which, however attractive, still lacks a demonstration of its
technical feasibility  and a sharp scientific justification for its
large cost.   Proposals for a
U.S.-sited Higgs factory have been discussed at Snowmass, including ILC
in the U.S. and other options.  However, we should not restrict
ourselves to a strictly regional approach. The world needs a coherent plan,
involving all of the regions active in particle physics, to carry out
this important physics program on a time scale relevant to the
students and postdocs now beginning their careers at the LHC. P5 must
help break the impasse by assigning the idea of an $\ee$ Higgs factory high
priority, initiating a global discussion of the technology choice and
cost sharing, and offering the option of siting the Higgs factory in the U.S.

\newsection{Linear vs. circular} 

Proposed Higgs factories come in two types---linear and circular.
Both types of accelerator  can meet the basic demands of the Higgs
factory physics case that we have presented here.   However, we feel
that the proposed linear colliders---in particular, the ILC---have
significant
advantages over the proposed circular colliders.   It is true that
circular $\ee$ collider designs can claim higher instantaneous luminosity
than established linear collider designs, at least for center of mass energies up
to 250~GeV.  However, the ILC offers important compensatory
advantages:
\begin{enumerate}
\item {\bf Beam polarization:}  $\ee$ colliders primarily enable studies of the
  electroweak interactions, which have order-1 parity violation. In the
  Standard Model, the left- and right-handed chiral components of
  fermions have different quantum numbers, signalling a different
  fundamental origin.  With $e^+$ and $e^-$ longitudinal beam polarization, we can
  selectively study the properties of these states.  This can
  compensate for smaller luminosity (by a factor of 2.5 in the Higgs
  boson studies), but more importantly it allows us to study more
  specific observables.
\item {\bf Design for precision:} Observation of a deviation from the Standard Model will require precise measurements with excellent control of systematic experimental uncertainties. Linear colliders offer a larger suite of tools for this purpose. Examples include the use of power-pulsing of the detector electronics between bunch trains to reduce cooling requirements, allowing a much reduced material budget in the tracker. The experiments can use larger magnetic fields without perturbing the beams, allowing high-resolution, compact detectors. The greater distance between beamline components at the interaction point allows detector coverage to much smaller angles. For control of systematics, electron and positron polarization can be used to produce relatively background-free and background-dominated data sets.
\item{\bf Higher energy:}  Though the luminosity of circular $\ee$
  colliders is high at low energy, it plummets at center of mass
  energies above 350~GeV.  The luminosity of a linear collider
  increases
  roughly linearly with energy, enabling energy upgrades to 500~GeV and
  beyond.   This is part of the ILC plan.  It allows us to measure the full suite of
  top quark electroweak form factors, the top quark-Higgs Yukawa
  coupling, and the cross section for double Higgs production, sensitive to
  the Higgs self-coupling.  These elements are essential parts of a
  complete precision measurement program for the Standard Model.
  \item{\bf Cost:} Proposed circular Higgs factories require tunnels
    of order 100~km in 
size. The estimated cost of the tunnel and civil engineering for the FCC is
comparable to the full estimated cost of the 250-GeV ILC.
 \end{enumerate}

 \newsection{Resources for P5}

Because the question of a global Higgs factory is central to the
mission of P5, the ILC community has created a detailed reference work
that covers all of the aspects of the ILC program and the program of
Higgs factories more generally.  This volume, 
the {\it ``International Linear Collider Report to Snowmass
2021''}~\cite{ILC2022izu}, 
is 350 pages long
and contains almost 800 references; it represents the
work and opinions of more than 500 authors. We do not expect all 
members of P5 to read the full report, but we expect that many
will find this volume useful both as an introduction to
Higgs factory physics and as a guide to the literature on this
subject.  In the next few paragraphs, we
review its contents.

Chapters 1--3 provide a general introduction and a description of the
ILC accelerator and physics organization.

Chapter 4 describes the ILC accelerator.  We claimed above that the
ILC is the most mature  Higgs factory proposal.  In this chapter, we
defend that statement through a detailed review of the ILC design and
the accelerator issues that it takes into account.  This chapter also
includes a discussion of the ILC cost, which has been thoroughly
reviewed as a part of the evaluation in Japan. It also reviews
strategies to make the
ILC sustainable (``Green ILC'').

Chapter 5 is an introduction to particle physics processes at Higgs
factory energies.  It gives perspective on the various Standard Model
processes and  background sources.  It also introduces the physics of
electron and positron beam polarization, a unique tool of
linear $e^+e^-$ colliders, covering both the precision measurement of
beam polarization and the  variety of its applications to the ILC
experimental program. 

Chapter 6 describes the detectors for ILC.   Two complete
designs, ILD and SiD, 
have been created and studied in detail at the full-simulation level.
This discussion brings up to date our current understanding of what is
possible for precision measurement in collider physics.
Ideas to improve the performance of Higgs factory measurements even further
are also discussed.  Simulation studies require also a complete
simulation framework, which is described in Chapter 7.   Projections of
physics capabilities quoted later in the book are obtained using 
full-simulation data and the described analysis toolkit, allowing us
to account for all anticipated sources of physics and
machine-induced backgrounds and systematic uncertainties.

Chapter 8 reviews in detail the physics simulation studies done
for the ILC at a center of mass energy of 250~GeV.
The experimental program here is already broad, covering not only Higgs boson
measurements but also the highest-precision  studies of the $W$ boson
couplings
and of perturbative QCD.    These, and the work described in Chapters~9 and
10, are the
most complete studies done for any proposed Higgs factory.  However,
since the $e^+e^-$ environment is relatively benign, their results can also
be used to estimate the capabilities of alternate proposals. 

Chapter 9 reviews the capability of the ILC for improved precision
measurements of the electroweak interactions.   The importance of beam
polarization is especially striking in precision electroweak
measurements.  We show in Chapter 9 that the measurement of
$\sin^2\theta_w$ using beam polarization can compensate a factor of
$10^3$ in luminosity, while at the same time offering better control
of systematic errors.

Chapter 10 reviews the physics simulation studies at the top quark
threshold and at 500~GeV,  covering studies of the Higgs boson in $WW$
fusion and the Higgs self-coupling measurement, 
measurements of $W$ boson and top quark interactions, the study of
fermion pair production, and the direct search
for new elementary particles.  

Chapter 11 describes the proposed fixed-target program of the ILC,
accessing dark sector particles by the use of high energy electron and
positron beams and exploring strong-field QED in extreme regimes
relevant to active galaxies and cosmic ray production.

Chapter 12 describes the use of Standard Model Effective Field Theory
to combine data from Higgs boson, $W$ boson, and electroweak
measurements.  This method gives the most powerful constraints from
experiment
 to compare to Standard Model predictions, including a
model-independent determination of the Higgs boson total width.  This 
chapter also provides a summary of our projections for the values of
the uncertainties in the measurement of Higgs boson couplings expected at
the ILC.

Chapters 13 and 14 discuss the theoretical interpretation of Higgs
factory measurements.   Chapter 13 describes how the Higgs factory
program sheds light on the major  large-scale questions of particle
physics. Chapter 14 discusses the comparison of Higgs factory
capabilities to the predictions of beyond-Standard-Model theories.
This chapter also estimates  the new particle  mass scales that can be
accessed by Higgs factory precision measurements.

Finally, Chapter 15 presents ideas for future colliders that could
make use of the ILC Laboratory after the ILC program ends. With new 
acceleration technologies,  the long,
straight tunnel needed for a linear collider could be the basis for a
program at much higher energy, reaching to 10~TeV and beyond.

We believe  that anyone who studies this volume will be convinced that
an $e^+e^-$ Higgs factory, especially one based on a linear collider,
provides a superb opportunity to advance our understanding of the most
important issues in particle physics.

And we expect that, armed with this understanding, P5 will create a
plan that will make this opportunity available to the current younger generation
of particle physicists.

\newsection{Acknowledgements}
We are grateful to Jenny List, Roman P\"oschl and Caterina Vernieri for their
useful comments.

\end{document}